\documentclass[twocolumn,reprint,amsmath,amssymb]{revtex4}

\usepackage{graphicx}% Include figure files
\usepackage{graphics}
\usepackage{dcolumn}% Align table columns on decimal point
\usepackage{bm}% bold math

\def\BEq{\begin{equation}}
\def\EEq{\end{equation}}
\def\BEqA{\begin{eqnarray}}
\def\EEqA{\end{eqnarray}}
\def\BEn{\begin{enumerate}}
\def\EEn{\end{enumerate}}
\def\BWT{\begin{widetext}}
\def\EWT{\end{widetext}}
\allowdisplaybreaks[1]

\begin{document}

%\preprint{APS/123-QED}

\title{Lamb's reconstruction of potentials and spatially localized scattering 
in nonrelativistic quantum mechanics}

\author{Andrei Galiautdinov$^1$}
 \affiliation{
$^1$Department of Physics and Astronomy, 
University of Georgia, Athens, GA 30602, USA}

\date{\today}% It is always \today, today,

\begin{abstract}

We formulate a simple condition for reconstructibility of a certain class of 
Hamiltonians with real potentials from the knowledge of their 
complex-valued eigenfunctions. This may be relevant to the question of 
preparability of quantum states raised by W.\ Lamb in his 1969 paper on 
operational interpretation of quantum mechanics. Of particular interest to 
engineering applications are: (a) an exotic case of an upside-down 
harmonic-oscillator-type potential (a variation of the inverted ``Mexican hat'' 
potential) whose square-integrable complex eigenfunction describes a localized 
scattering state similar to the ``bound state in the continuum'' of von Neumann 
and Wigner, and (b) a spatially confined scattering state of a particle moving in 
an infinite well with the properly shaped potential bottom.

\end{abstract}

%\pacs{03.75.Dg}

\maketitle

%\tableofcontents
%\newpage
%\clearpage
%\vskip50pt

\allowdisplaybreaks[1]

\section{Introduction}

In 1969, Willis Lamb, Jr., published a remarkable paper \cite{Lamb1969} 
(also \cite{Lamb1979}) on the operational foundations of nonrelativistic 
quantum mechanics, in which he proposed, among other things, an experimental 
procedure for preparing an arbitrary quantum state, $\psi(x)=R(x)e^{iS(x)}$, of 
a particle moving on the $x$-axis. The main assumption that had to be made in 
Ref.\ \cite{Lamb1969} was the availability of an arbitrary potential field, $V(x,t)$, 
with which to manipulate the corresponding quantum system. The preparation 
procedure consisted of the following four steps: 
\begin{enumerate}
\item calculating, on the basis of $\psi(x)$, of a certain physically realizable 
potential, $V_1(x)$, 
\item experimentally setting up the calculated $V_1(x)$, 
\item catching the particle in one of the eigenstates, $\psi_1(x)\equiv R(x)$, 
of the resulting Hamiltonian, and, once the particle was caught,  
\item subjecting the particle to an additional pulse perturbation of the form 
$-S(x)\delta(t)$, which, as seen from the time-dependent Schr\"{o}dinger equation,
\BEq
i\frac{\partial \psi(x,t)}{\partial t}=-\frac{1}{2}\frac{\partial^2 \psi(x,t)}{\partial x^2}
-S(x)\delta(t)\psi(x,t),
\EEq  
would immediately bring the particle to the desired state $\psi(x)$.
\end{enumerate}

The final step involving the pulsed perturbation complicated the preparation 
procedure. Lamb resorted to that step because the method by which $V_1(x)$ 
was calculated (see Sec.\ \ref{sec:solution}), when applied directly to the complex 
$\psi(x)$, resulted in a complex and thus unphysical potential (at least, as far 
as non-$PT$-symmetric case is concerned \cite{Bender1998, Bender2005}). 
As a result, if one wanted to shorten the procedure by dropping the final step, 
one had to restrict consideration to real eigenfunctions only. Such a restriction 
limits experimentalist's ability to control the quantum system. It is therefore 
natural to attempt to bypass that restriction by expanding the set of experimentally 
available wave functions associated with real, physically realizable (at least, in 
principle), potentials. In the following sections we find the conditions that must be 
imposed on a complex wave function in order for it to be an eigenfunction of a 
Hamiltonian with real potential. The set of the corresponding potentials 
turns out to be quite limited and consists of rather exotic field configurations that 
are not typically encountered in a laboratory.

\section{Formulation of the problem}

In one dimension, the task is: Given a complex wave function, $\psi(x)$, reconstruct 
the Hamiltonian, $H$, for which that function is an eigenfunction.

The Hamiltonian is assumed to be of the form
\BEq
H = -\frac{1}{2}\frac{d^2}{dx^2} + V(x),
\EEq
with the corresponding time-independent Schr\"{o}dinger's equation being
\BEq
\label{eq:2}
-\frac{1}{2}\frac{d^2\psi(x)}{dx^2} + V(x)\psi(x) = E\psi(x).
\EEq
The given eigenfunction is assumed to be
\BEq
\label{eq:3}
\psi(x) = R(x)e^{iS(x)},
\EEq
with real $R(x)$ and $S(x)$, as proposed in Ref.\ \cite{Lamb1969}. 
To avoid problems with singular solutions, we restrict consideration to 
nodeless $R(x)$ only. The goal is to find a {\it real} potential $V(x)$ 
that satisfies both (\ref{eq:2}) and (\ref{eq:3}).

\section{Reconstruction Procedure}

\label{sec:solution}

When $\psi(x)$ is real, the procedure due to Lamb \cite{Lamb1969} 
(and also Berezin, as recounted in \cite{Turbiner2016}) is to simply 
invert (\ref{eq:2}) and get 
\BEq
\label{eq:2-invertedSimple}
V(x)= E+\frac{1}{2}\frac{\psi''(x)}{\psi(x)},
\EEq
where $E$ can be chosen arbitrarily (the prime $'$ indicates differentiation 
with respect to $x$). When $\psi(x)$ is complex, this procedure fails (we do 
not get real $V(x)$), unless specific ``reconstructability'' conditions are satisfied. 
Our goal is to determine those conditions.

 To determine the reconstructability conditions, we first notice that due to its 
linearity the Schr\"{o}dinger equation (\ref{eq:2}) is separately satisfied by 
each of the real and imaginary parts of $\psi(x)$. Thus, writing
\BEq
\psi(x) = a(x)+ib(x),
\EEq
with real $a(x)$ and $b(x)$, we get
\begin{align}
\label{eq:2a}
-\frac{1}{2}\frac{d^2 a(x)}{dx^2} + V(x)a(x) &= Ea(x), 
\\
\label{eq:2b}
-\frac{1}{2}\frac{d^2 b(x)}{dx^2} + V(x)b(x) &= Eb(x).
\end{align}
Inverting each of these equations automatically gives two real potentials, 
\begin{align}
\label{eq:3a}
V_a(x)&= E+\frac{1}{2}\frac{a''(x)}{a(x)}, 
\\
\label{eq:3b}
V_b(x)&= E+\frac{1}{2}\frac{b''(x)}{b(x)},
\end{align}
and, since in both cases we must get the same $V(x)$, the sought 
reconstructability condition reads
\BEq
\label{eq:reconstrucatbilityCondition}
\frac{a''(x)}{a(x)}= \frac{b''(x)}{b(x)}.
\EEq
To recast (\ref{eq:reconstrucatbilityCondition}) in terms of $R(x)$ 
and $S(x)$, we write
\begin{align}
\label{eq:4}
\psi(x) &= R(x)\cos S(x) + i R(x)\sin S(x), 
\end{align}
and find
\begin{align}
\label{eq:5a}
\frac{a''}{a} 
&\equiv \frac{(R\cos S)''}{R\cos S}
= \frac{R''-R(S')^2}{R}-\frac{2R'S'+RS''}{R}\tan S, 
\\
\label{eq:5b}
\frac{b''}{b} 
&\equiv \frac{(R\sin S)''}{R\sin S}
= \frac{R''-R(S')^2}{R}+\frac{2R'S'+RS''}{R}\frac{1}{\tan S}.
\end{align}
Eq.\ (\ref{eq:reconstrucatbilityCondition}) demands that in (\ref{eq:5a}) 
and (\ref{eq:5b}) we  set
\BEq
\label{eq:reconstrucatbilityConditionNEW}
2R'S'+RS''=0,
\EEq
which gives
\BEq
\label{eq:6}
S(x) = C \int^{x} \frac{ds}{R^2(s)},
\EEq
where $C$ is an arbitrary real constant.
Thus, for a given $R(x)$ in (\ref{eq:3}), we choose $S(x)$ in accordance with 
(\ref{eq:6}), and using (\ref{eq:5a}) get the real potential  
(cf.\ \cite{vonNeumann1929, Stillinger1975, Khelashvili1996, Petrovic2002}), 
\BEq
\label{eq:V(x)}
V(x)= E + \frac{R''}{2R}-\frac{C^2}{2R^4}.
\EEq
This results in the Hamiltonian,
\BEq
H = -\frac{1}{2}\frac{d^2}{dx^2} + \frac{R''}{2R}-\frac{C^2}{2R^4} + E,
\EEq
whose complex eigenfunction of energy $E$ is given by
\BEq
\label{eq:7}
\psi(x) = R(x)\exp\left\{iC \int^{x} \frac{ds}{R^2(s)}\right\},
\EEq 
where the lower limit of integration may be chosen arbitrarily; different 
choices will result in unimportant phase shifts. For nodeless $R(x)$, the 
function $\psi(x)$ satisfies all the usual properties of a ``legitimate'' wave 
function \cite{LL1977}: it is single-valued and continuous on the entire $x$-axis. 
Additionally, if $R(x)$ is square-integrable, then so is the 
$\psi(x)$. The characteristic feature of all the found wave functions is that the 
associated current densities,
\begin{align}
\label{jx}
j_x(x) = R^2(x) S'(x) = C,
\end{align} 
are constant throughout their respective domains of definition.

In higher dimensions, the Schr\"{o}dinger equation reads (here, ${\bf r}$ is 
particle's position vector, $\nabla$ is the nabla operator)
\BEq
\label{eq:2-3D}
-\frac{1}{2}\nabla^2\psi({\bf r}) + V({\bf r})\psi({\bf r}) = E\psi({\bf r}),
\EEq
with the analogues of Eqs.\ (\ref{eq:reconstrucatbilityCondition}) and 
(\ref{eq:reconstrucatbilityConditionNEW}) being, respectively,
\BEq
\frac{\nabla^2 a}{a}= \frac{\nabla^2 b}{b},
\EEq
and
\BEq
2\nabla R \cdot \nabla S+R\nabla^2S =0.
\EEq
For example, in the spherically symmetric case in three dimensions, we get
\BEq
2\frac{dR}{dr}\frac{dS}{dr}+\frac{R}{r^2}\frac{d}{dr}\left(r^2\frac{dS}{dr}\right) =0,
\EEq
which gives,
\BEq
\label{eq:6-high}
S(r) = C \int^{r} \frac{ds}{s^2R^2(s)},
\EEq
and, thus,
\BEq
\label{eq:V(x)-high}
V(r)= E + \frac{1}{2Rr^2}\frac{d}{dr}\left(r^2\frac{dR}{dr}\right) 
-\frac{C^2}{2r^4R^4},
\EEq
with the corresponding eigenfunction being
\BEq
\label{eq:7-1}
\psi(r) = R(r)\exp\left\{iC \int^{r} \frac{ds}{s^2R^2(s)}\right\}.
\EEq 

%%%%%%%%% BEGIN FIG. 1
\begin{figure*}[!ht]
\includegraphics[angle=0,width=1.00\linewidth]{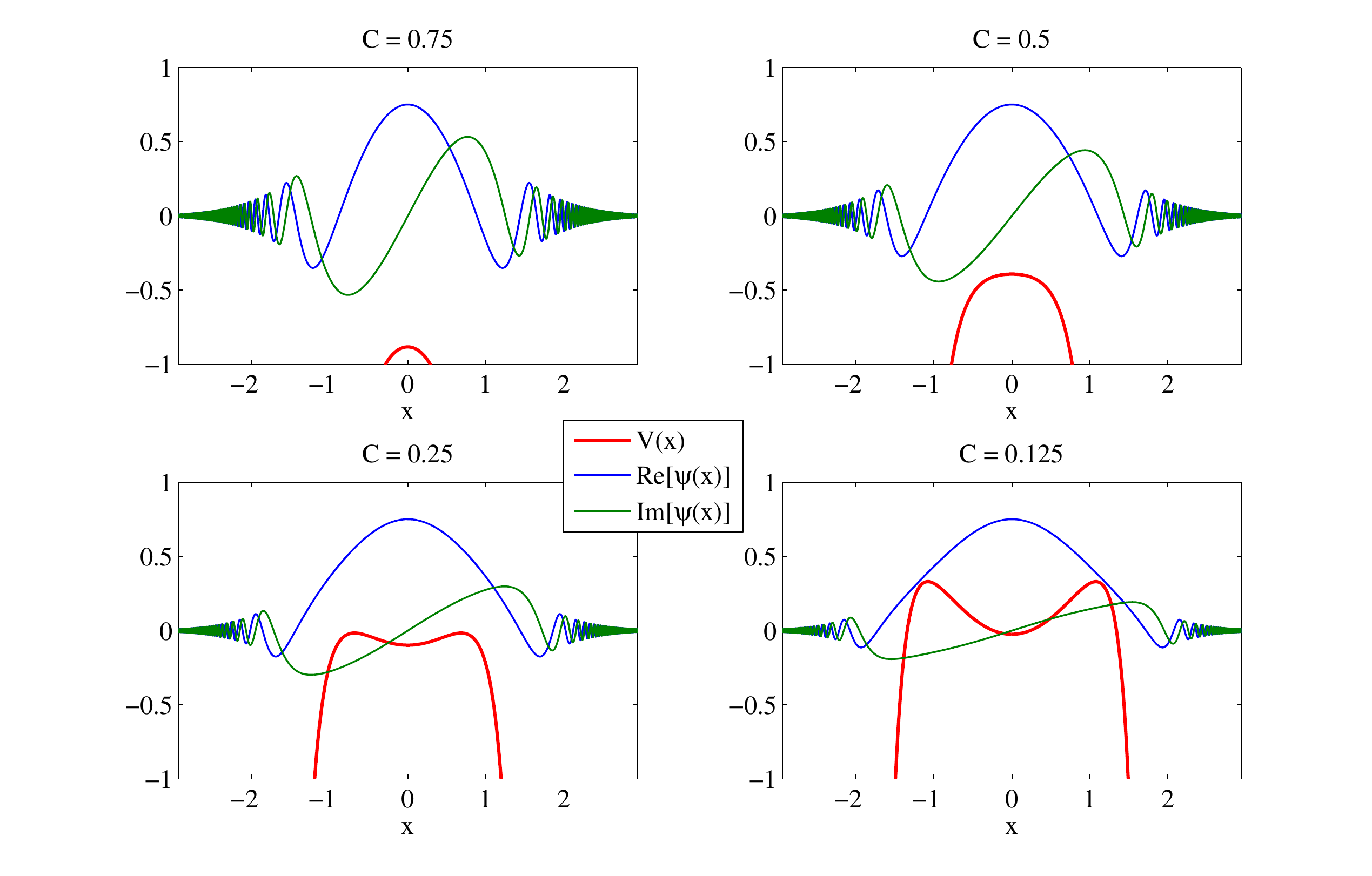}
\caption{ \label{fig:1} 
(Figure online.) Deformed harmonic-oscillator case of Sec.\ \protect\ref{sec:HO-0}, Eqs.\ 
(\protect\ref{eq:8}) and (\protect\ref{eq:9}): Plots of the reconstructed potential, $V(x)$, as 
well as the real and imaginary parts of its complex nodeless eigenfunction, $\psi(x)$, 
for several values of $C$. The lower limit of integration in (\protect\ref{eq:8}) was set to zero. 
When $C=0$, we recover the standard harmonic oscillator result.}
\end{figure*}
%%%%%%%%% END FIG. 1

\section{Example: Free particle in one dimension}
\label{sec:FP}

The simplest example is provided by a wave function with prefactor
\BEq
R(x)=1,
\EEq
or,
\BEq
\label{eq:FP1}
\psi(x) = \exp\left\{iC \int_0^{x} ds\right\} =  e^{iC x}.
\EEq 
The corresponding potential is immediately found to be
\BEq
\label{eq:FP2}
V(x) = 0,
\EEq
where we set $E=C^2/2$. This is the usual expression for particle's kinetic 
energy in terms of its momentum $C$. Being trivial, this result serves as a 
useful consistency check for our approach.

\section{Example: One-dimensional Harmonic oscillator}
\label{sec:HO-0}

Here we consider a localized wave function with prefactor
\BEq
R(x)=\left(\frac{1}{\pi}\right)^{1/4}e^{-x^2/2},
\EEq
so that
\BEq
\label{eq:8}
\psi(x) 
= \left(\frac{1}{\pi}\right)^{1/4}
e^{-x^2/2}\exp\left\{iC \sqrt{\pi} \int_0^{x} e^{s^2}ds\right\}.
\EEq 
The corresponding potential, according to (\ref{eq:V(x)}), is
\BEq
\label{eq:9}
V(x) = \frac{x^2}{2}-\frac{C^2 \pi e^{2x^2}}{2},
\EEq
where we set $E=1/2$. 
When $C=0$, we get the standard harmonic oscillator result.
When $C\neq 0$, the harmonic oscillator potential deforms into an 
upside-down shape (a variation of inverted ``Mexican hat'' potential), 
as shown in Fig.\ \ref{fig:1}, with the phase of the wave function rapidly 
increasing as $|x|\rightarrow \infty$. Thus, what we are dealing with here 
is a singular Sturm-Liouville problem with a negatively diverging potential, 
$V(x)\rightarrow -\infty$, on both ends of the corresponding boundary-value 
interval, $|x|\rightarrow \infty$. This is reminiscent of the situation encountered 
in  $PT$-symmetric quantum mechanics \cite{Bender1998}, where some 
Hamiltonians, such as, e.\ g.,
\BEq
H_{PT} = {p^2} + x^2 (ix)^N, \quad N = 2,
\EEq
often come with upside-down potentials. The difference between the two 
approaches is that in our case the resulting Hamiltonian is Hermitian (in the 
standard Dirac's sense) and does not require any redefinition of the usual 
scalar product on the Hilbert space of states.

The above result is unusual in that the probability current corresponding to 
(\ref{eq:8}) is,
according to Sec.\ \ref{sec:solution}, constant throughout the entire $x$-axis,
\begin{align}
\label{jx-HO}
j_x = R^2 S' = C,
\end{align} 
while the probability density,
 \begin{align}
\rho = R^2 = \frac{e^{-x^2}}{\sqrt{\pi}},
\end{align}
is exponentially decreasing with the square of the distance from the origin.
Eq.\ (\ref{jx-HO}) is a signature of a scattering state. It shows that $\psi(x)$ 
represents a particle impinging from the left (if $C>0$) on the corresponding 
potential barrier and then accelerating away to (positive) infinity, while spending 
most of its ``life'' being localized on the barrier's top (compare with the 
above-barrier localized states in double-well potentials considered in 
Refs.\ \cite{Dutta1992, Dutta1996}; also see 
\cite{vonNeumann1929, Stillinger1975, Khelashvili1996, Petrovic2002} 
and \cite{Simon1967, Arai1999} for the so-called ``bound states in the 
continuum'' of von Neumann and Wigner). 

A natural question arises whether the found potential (\ref{eq:9}) is physically 
realistic. Superficially, the answer seems to be ``no,'' since it is well-known that 
a classical particle subjected to a much milder potential $V(x) = -|x|^{\alpha}$ 
would reach infinity in finite time for any $\alpha > 2$ 
(see, e.\ g., \cite{ReedSimonII}, \cite{Simon2000}), and out potential is much more 
singular than that. However, one should keep in mind that, from the physical 
standpoint, even the usual harmonic oscillator potential cannot be regarded as 
realistic, for there does not exist a system in nature for which $V(x)=x^2/2$ 
for all $x$. Nevertheless, the harmonic oscillator often serves as a useful 
approximation to actual, physically realizable situations: its first few states 
describe various quantum systems quite well. We expect something similar 
to take place with our singular potentials too. For example, as a useful approximation, 
we can construct a potential that has the form (\ref{eq:9}) on a finite interval, 
$|x| \leq a <+\infty$, and adjust the system parameters so that the real part 
of $\psi$ would vanish at $|x|=a$. The resulting configuration would then be 
indistinguishable from the one involving a particle confined to an infinite well 
with the walls at $x = \pm a$. Of course, such scenario is not as dramatic as 
the localized scattering on an infinite line. However, it may be possible to create 
some finite, physically realistic, periodic configurations (say, motion on a circle, 
such as in the case of superconducting Josephson phase qubits \cite{DM2004}), 
whose quantum states exhibit properties similar to those of the localized scattering 
states considered above. 

%%%%%%%%% BEGIN FIG. 2
\begin{figure*}[!t]
\includegraphics[angle=0,width=1.00\linewidth]{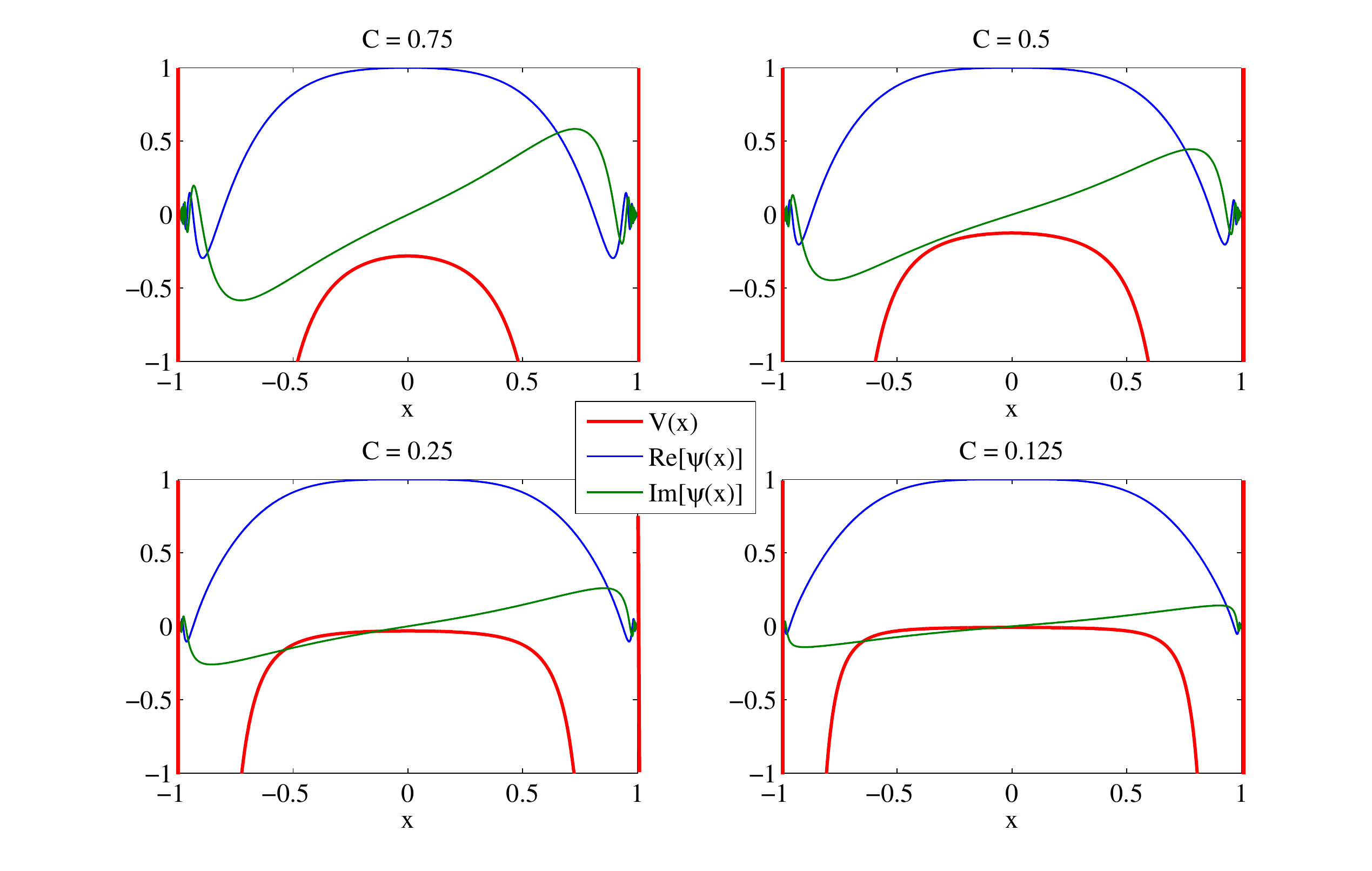}
\caption{ \label{fig:2} 
(Figure online.) Deformed infinite square well case of Sec.\ \protect\ref{sec:ISW}, Eqs.\ 
(\protect\ref{eq:10}) and (\protect\ref{eq:11}): Plots of the reconstructed potential, $V(x)$, as 
well as the real and imaginary parts of its nodeless complex eigenfunction, $\psi(x)$, 
for several values of $C$. The lower limit of integration in (\protect\ref{eq:7}) was set to zero.}
\end{figure*}
%%%%%%%%% END FIG. 2
%%%%%%%%% BEGIN FIG. 3
\begin{figure*}[!t]
\includegraphics[angle=0,width=1.00\linewidth]{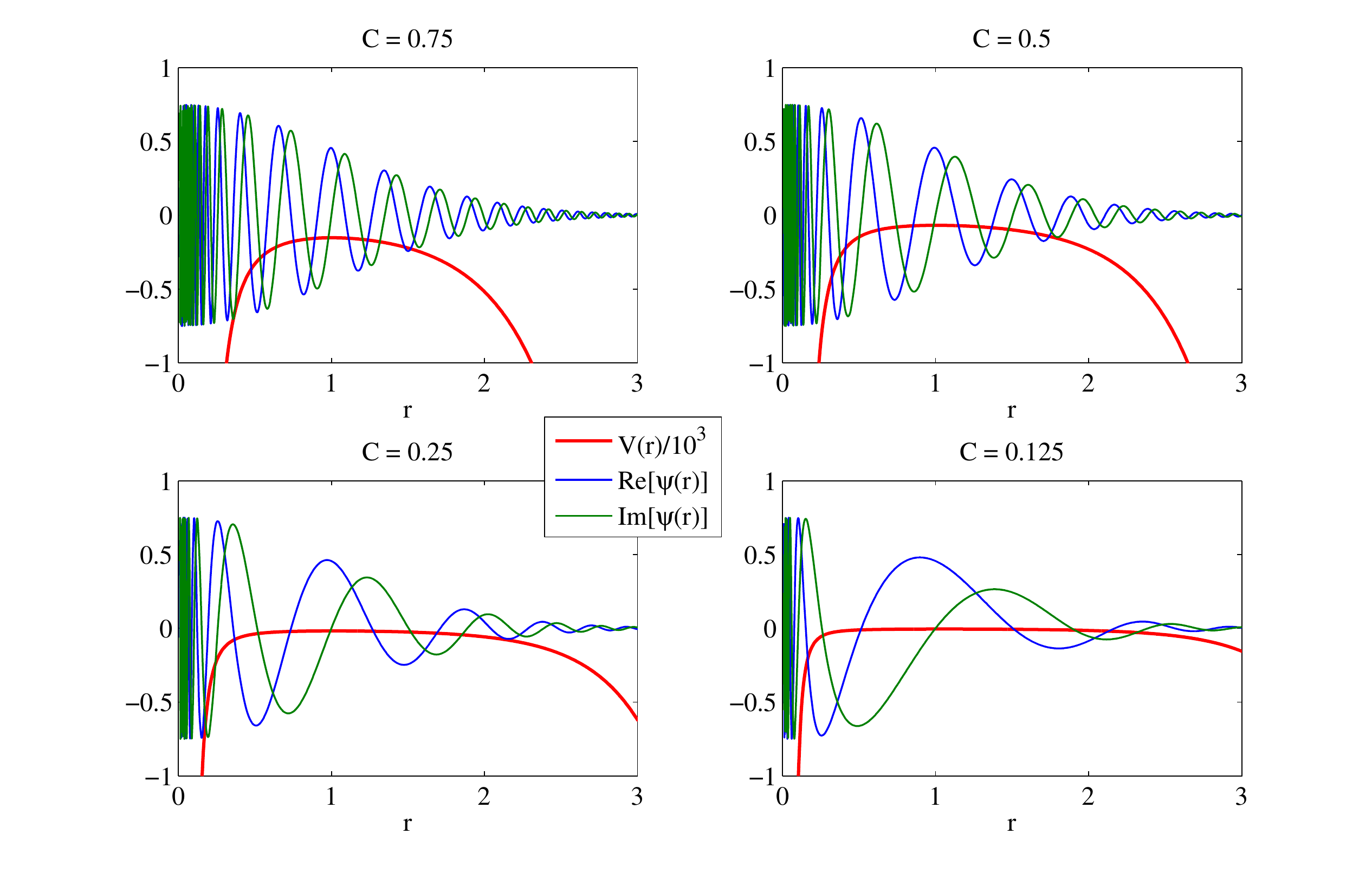}
\caption{ \label{fig:3} 
(Figure online.) Deformed hydrogen atom potential of Sec.\ \protect\ref{sec:HA}, Eqs.\ 
(\protect\ref{eq:psi-HA}) and (\protect\ref{eq:V(x)-HA}): Plots of the reconstructed potential, 
$V(x)$, as well as the real and imaginary parts of its nodeless complex 
eigenfunction, $\psi(x)$, for several values of $C$. The lower limit of integration 
in (\protect\ref{eq:psi-HA}) was set to 1.}
\end{figure*}
%%%%%%%%% END FIG. 3

\section{Example: One-dimensional infinite square well}
\label{sec:ISW}

In this example, the prefactor is 
\BEq
R(x)=\cos\left(\frac{\pi x}{2}\right), \quad |x|< 1,
\EEq
which results in the eigenfunction
\BEq
\label{eq:10}
\psi(x) = \cos\left(\frac{\pi x}{2}\right)\exp\left\{iC\frac{2}{\pi} 
\tan\left(\frac{\pi x}{2}\right)\right\}, 
\quad 
|x|< 1.
\EEq 
The corresponding potential, according to (\ref{eq:V(x)}), is
\BEq
\label{eq:11}
V(x) = -\frac{C^2}{2\cos^4\left(\frac{\pi x}{2}\right)}, \quad 
|x|< 1,
\EEq
where we set $E=\pi^2/8$. 
When $C=0$, we recover the standard result for an infinite square well.
When $C\neq 0$, the potential bottom assumes an upside-down shape, 
as shown in Fig.\ \ref{fig:2}, with the phase of the wave function rapidly 
increasing as one approaches the infinite walls at $|x|= 1$. 

The probability current corresponding to (\ref{eq:10}) is constant throughout 
the interval
$|x|\leq 1$,
\begin{align}
\label{jx-ISW}
j_x = C,
\end{align} 
while the probability density varies with position as
 \begin{align}
\rho = \cos^2\left(\frac{\pi x}{2}\right).
\end{align}
In this case we have a scattering state confined in a box. Thinking classically, 
the particle always moves in one direction, never encountering the potential 
walls. This example complements rather nicely the remarks made at the end of 
the previous section.

\section{Example: Hydrogen atom}
\label{sec:HA}

Here we consider spherically symmetric prefactor,
\BEq
R(r)=\frac{1}{\pi^{1/2}}e^{-r},
\EEq
which gives the wave function,
\BEq
\label{eq:psi-HA}
\psi(r) 
= \frac{1}{\pi^{1/2}}e^{-r}\exp\left\{iC \pi \int^{r}_{1} 
\frac{e^{2s}}{s^2}ds\right\},
\EEq 
where we have conveniently set the lower limit of integration to 1 to avoid 
the singularity at the origin.
The corresponding potential, according to (\ref{eq:V(x)-high}), is 
(see Fig.\ \ref{fig:3})
\BEq
\label{eq:V(x)-HA}
V(r)= -\frac{1}{r} - \frac{C^2\pi^2 e^{4r}}{2r^4},
\EEq
where we have chosen $E=-1/2$. In this example, setting $C$ to zero 
reproduces the familiar ground state and Coulomb potential of the hydrogen atom.

\section{Conclusions}

In summary, we have given a sufficient condition for Lamb's reconstructability 
of certain real-valued potentials from the knowledge of their respective complex 
eigenfunctions. The found potentials are rather exotic and not typically encountered 
in nature. Nevertheless, as a matter of principle, our results could still be useful for 
designing new experimental techniques for quantum state preparation, manipulation, 
and control, as well as for the development of novel quantum electronic devices 
\cite{Vladimirova1999, Ordonez2006, Nakamura2007, Longhi2007, Moiseyev2009, Hsueh2010, Prodanovic2013}.

%\begin{acknowledgments}
%\end{acknowledgments}

\end{document}